\documentclass[12pt]{iopart}

\expandafter\let\csname equation*\endcsname\relax

\expandafter\let\csname endequation*\endcsname\relax

\usepackage{amsmath}
\usepackage{amsfonts}
\usepackage{iopams}
\usepackage{cite}
\usepackage[colorlinks=true, citecolor=blue]{hyperref}
\usepackage{graphicx}
\usepackage{hyperref}
\usepackage{graphics}
\usepackage{amssymb}
\usepackage{color}
\usepackage{bm}
\usepackage{float}
\usepackage[utf8x]{inputenc}

\begin{document}

\title[Quantum speed limit time for moving qubit inside a leaky cavity]{Quantum speed limit time for moving qubit inside a leaky cavity}
\author{Maryam Hadipour$^{1}$,  Soroush Haseli$^{2}$\footnote{
Corresponding author}, Hazhir Dolatkhah$^{3,4}$, and Saeed Haddadi$^{5,6}$}

\address{
$^{1}$ Department of Physics, Ahwaz Branch, Islamic Azad University, Ahvaz, Iran\\
$^{2}$ Faculty of Physics, Urmia University of Technology, Urmia, Iran\\
$^{3}$ RCQI, Institute of physics, Slovak Academy of Science, Dubravska Cesta $9$, $84511$ Bratislava, Slovakia\\
$^{4}$ Department of Physics, University of Kurdistan, P.O.Box 66177-15175, Sanandaj, Iran\\
$^{5}$ Faculty of Physics, Semnan University, P.O.Box 35195-363, Semnan, Iran\\
$^{6}$ Saeed’s Quantum Information Group, P.O.Box 19395-0560, Tehran, Iran
}
\ead{soroush.haseli@uut.ac.ir}

\begin{indented}
\item[]
\end{indented}

\begin{abstract}
The minimum time required for a quantum system to evolve from an arbitrary initial state to its orthogonal state is known as the quantum speed limit (QSL) time. In this work, we consider the model in which a single qubit moves inside a leaky cavity and then we study the QSL time for this model. Notably, we show that for both weak and strong coupling regimes, the QSL time increases with increasing the velocity of the qubit inside the leaky cavity. Moreover, it is observed that by increasing qubit velocity, the speed of the evolution tends to a constant value and the system becomes more stable.
\end{abstract}

%

%
%
%
%

\section{Introduction}\label{intro}
Quantum mechanics, as a fundamental law in nature, sets a bound on the speed of evolution of a quantum systems. This bound utilizes in various topics of quantum theory, including quantum communication \cite{1}, investigation of exact bounds in quantum metrology \cite{2}, computational bounds of physical systems \cite{3} and quantum optimal control algorithms \cite{4}. The shortest time a system needs to change from an initial state to its orthogonal state is called the quantum speed limit (QSL) time. This time has been studied for both closed and open quantum systems. For closed systems, geometric measures such as Bures angle and relative purity are used to define the bound of QSL time  \cite{5,6,7,8,9,10,11,12}. The first bound of QSL time  for closed quantum systems  has been introduced by Mandelstam and Tamm as \cite{11}
\begin{equation}\label{MT}
\tau \geq \tau_{Q S L}=\frac{\pi \hbar}{2 \Delta E},
\end{equation}
where $\Delta E =\sqrt{ \langle \hat{H}^2 \rangle  - \langle \hat{H}\rangle^2}  $ is the variance of energy of initial state and $\hat{H}$ shows the time-independent Hamiltonian. The bound in Eq. (\ref{MT}) is known as MT bound. Another bound of QSL time for closed quantum systems has been introduced by Margolus and Levitin as \cite{12}
\begin{equation}\label{ML}
\tau \geq \tau_{Q S L}=\frac{\pi \hbar}{2 E},
\end{equation}
where $E= \langle \hat{H} \rangle$. The bound in Eq. (\ref{ML}) is known as ML bound. In Ref. \cite{7}, Giovannetti et al. combine the MT and ML bounds for closed quantum systems and introduce a unified bound for QSL time as
\begin{equation}
\tau \geq \tau_{Q S L}=\max \left\{\frac{\pi \hbar}{2 \Delta E}, \frac{\pi \hbar}{2 E}\right\}.
\end{equation}
In the real world of quantum mechanics we are dealing with open quantum systems, so the study of these systems is of particular importance \cite{13,14,15}. So far, a lot of works have been done on QSL time for open quantum systems \cite{16,17,18,19,20,21,22,23,24,25,26,27,28,29,30,31,32,33,34,35,36,37,38,39,40,41}. It has been shown that the bound of QSL time for open quantum systems can be defined by extending the MT and ML bounds for these systems \cite{7,8}. In Ref. \cite{28}, Zhang et al. have introduced the bound for open quantum systems based on relative purity for arbitrary initial state. They have shown that QSL time is depend on quantum coherence. In this work, we will use Zhang et al. bound, which is presented in Sec. \ref{quantum}. The motivation for using this bound here is that this bound can be used for any desired initial state (pure or mixed).

The scheme of recent experiments in quantum information theory is based on the control of a qubit within a cavity. In practice, it is difficult to achieve the static state of a qubit in a cavity. Many studies have been done about the moving qubits in Markovian and non-Markovian environments \cite{42,43,44,45,46,47}. Markovian evolution is a memory-less evolution in which information is monotonically leaked from the system to the environment and there is no flow back of information from the system to the environment. While in non-Markovian evolution, the information flow back from the environment to system.
The protection of initial entanglement by moving two qubits in a non-Markovian environment has been studied in Ref. \cite{43}. In Ref. \cite{42}, the effects of qubit velocity within the leaky cavity on quantum coherence protection have been studied.  The dependence of QSL time on quantum coherence motivates us to study the effect of qubit velocity in a leaky cavity on QSL time \cite{48}. The motivation for choosing this model is that in recent works it has been shown that by increasing the qubit velocity inside the leaky cavity, the coherence of the qubit state is further protected. Therefore, we expect that as the qubit velocity increases within the leaky cavity, the QSL time will increase and evolution will be slower.

The work is organized as follow. In Sec. \ref{model}, the model is introduced. In Sec. \ref{quantum}, the QSL time for moving qubit is investigated.  Finally, we provide a conclusion in Sec. \ref{con}

\begin{figure}[H]
  \centering
  \includegraphics[width=0.70\textwidth]{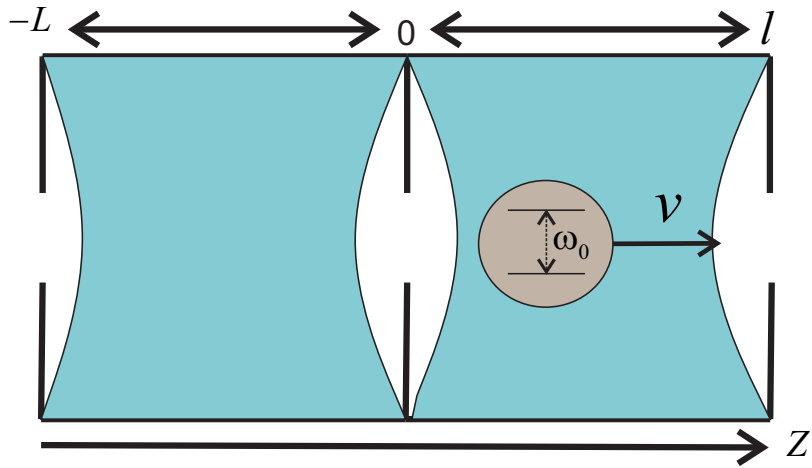}
\caption{Schematic diagram
of the model where the qubit
moves inside a leaky cavity with a constant velocity of $v$.}
\label{figure1}
\end{figure}

\section{Moving qubit inside a leaky cavity}\label{model}
Here, let us consider a model that includes a single-qubit system and a structured environment. The structure of the environment consists of two mirrors at $z=-L$ and $z=l$ and a mirror with partial reflection at $z=0$. It can be said that the environment consists of two consecutive cavities in intervals $(-L,0)$ and $(0,l)$. The structure of the model is shown in Fig. \ref{figure1}. Any classical electromagnetic field $E(z,t)$ in $(-L,l)$ can be expanded as \cite{42,43}
\begin{equation}
E(z, t)=\sum_{k} E_{k}(t) U_{k}(z)+E_{k}^{*}(t) U_{k}^{*}(z),
\end{equation}
where $U_k(z)$'s are exact monochromatic modes at frequency $\omega_k$ and $E_k(t)$ is the amplitude in the $k$-th mode. Herein, it is assumed that the electromagnetic field is polarized along the $x$-axis. To meet the boundary conditions in mirrors, the mode functions should be
\begin{equation}
U_{k}(z)=\left\{\begin{array}{ll}
\varepsilon_{k} \sin k(z+L) & z<0 \\
J_{k} \sin k(z-L) & z>0
\end{array}\right.
\end{equation}
where $\varepsilon_{k}$ takes the value  $\pm1$ going from each mode to the subsequent one and for a good cavity, we have
\begin{equation}
J_{k}=\frac{\sqrt{c \lambda^{2} / l}}{\sqrt{\left(\omega_{k}^{2}-\omega_{n}\right)^{2}+\lambda^{2}}},
\end{equation}
where $\omega_n=n \pi c /l$ is the frequency of the quasi mode and $\lambda$ represents the damping of the cavity in $(0, l)$. $\lambda$  determines the photon leakage through the cavity mirrors and also determines the spectral width of the coupling. Here, we assume that the qubit interacts with the second cavity located in the range $(0,l)$. The qubit also moves along the $z$-axis at a constant speed $v$.  The qubit interacts with the cavity modes as it moves inside the cavity. Considering the dipole and rotating-wave approximation, the Hamiltonian of the system is defined as follows
\begin{equation}
\begin{aligned}
\hat{H} &=\omega_{0}|0\rangle_{s}\langle 0|+\sum_{k} \omega_{k} a_{k}^{\dagger} a_{k} \\
&=\sum_{k} f_{k}(z)\left(g_{k}|1\rangle_{s}\left\langle 0\left|a_{k}+g_{k}^{*}\right| 0\right\rangle_{s}\langle 1| a_{k}^{\dagger}\right),
\end{aligned}
\end{equation}
where $\vert 0 \rangle_{s}$ ($\vert 1 \rangle_{s}$) is excited (ground) state of the system, $\omega_0$ denotes the transition frequency of the qubit, $a_k^{\dag}$ ($a_k$) is creation (annihilation) operator for the $k$-th cavity mode with frequency $\omega_k$, and $g_k$ specifies the coupling between qubit and cavity. The shape function describing the motion of the qubit along the $z$-axis is given by \cite{49,50,51}
\begin{equation}
f_{k}(z)=f_{k}(v t)=\sin [k(z-l)]=\sin \left[\omega_{k}(\beta t-\tau)\right],
\end{equation}
where $\beta = v/c$ and $\tau =l/c$. It has been shown that the parameter $\beta$ can be expressed as $\beta=v / c=(x) \times 10^{-9}$ \cite{42}. It is equivalent to $v = 0.3(x)$ for $^{85}Rb$ Rydberg microwave qubit. Let us assume that the general initial state of the total system is given by
\begin{equation}\label{initial}
|\psi(0)\rangle=\left(c_{1}|0\rangle_{s}+c_{2}|1\rangle_{s}\right) \otimes|0\rangle_{c},
\end{equation}
where we have $\vert c_1 \vert^{2}+\vert c_2 \vert^{2} =1$. The initial state of the cavity is in a vacuum state $|0\rangle_{c}$. After a while, the system state changes to the following form at time $t$
\begin{equation}
|\psi(t)\rangle=c_{1} A(t)|0\rangle_{s}|0\rangle_{c}+c_{2}|1\rangle|0\rangle_{c}+\sum_{k} B_{k}(t)|1\rangle_{s}\left|1_{k}\right\rangle_{c},
\end{equation}
in which $\left|1_{k}\right\rangle_{c}$ is the cavity state that includes one photon in the $k$-th mode. Using schr\"{o}dinger equation, differential equations for the probability amplitudes $A(t)$ and $B_k(t)$ can be obtained as follows
\begin{eqnarray}
i \dot{A}(t)&=&\omega_0 A(t) +  \sum_k g_k J_k f_k(vt)B_k(t), \label{ss1}\\
i \dot{B}_k(t)&=&\omega_k B_k(t) +g^{*}_k J_k f_k(vt)A(t). \label{ss2}
\end{eqnarray}
By solving Eq. (\ref{ss2}) and putting its results in Eq. (\ref{ss1}), we will have
\begin{equation}\label{ss3}\small \small
\dot{A}(t)+i \omega_{0} A(t)=-\int_{0}^{t} d t^{\prime} A\left(t^{\prime}\right) \sum_{k}\left|g_{k}\right|^{2} J_{k}^{2} f_{k}(v t) f_{k}\left(v t^{\prime}\right) e^{-i \omega_{k}\left(t-t^{\prime}\right)}.
\end{equation}
By defining the probability amplitude as $A\left(t^{\prime}\right)=\tilde{A}\left(t^{\prime}\right) e^{i \omega_{0} t^{\prime}}$ and putting it into Eq. (\ref{ss3}), we have
\begin{equation}\label{ss11}
\dot{\tilde{A}}+\int_{0}^{t}dt^{\prime}F(t,t^{\prime})\tilde{A}(t^{\prime})=0,
\end{equation}
where the memory kernel $F(t,t^{\prime})$ is defined as follows
\begin{equation}
F(t,t^{\prime})=\sum_k \vert g_k \vert^{2}J_k^{2}f_k(vt)f_k(vt^{\prime})e^{-i(\omega_k-\omega_0)(t-t^{\prime})}.
\end{equation}
This memory kernel in the continuum limit has the following form
\begin{equation}
F(t,t^{\prime})=\int_0^{\infty} J(\omega_k)f_k(t,t^{\prime})e^{-i(\omega_k-\omega_0)(t-t^{\prime})}d\omega_k,
\end{equation}
where $f_k(t,t^{\prime})=\sin[\omega_k(\beta t -\tau)]\sin[\omega_k(\beta t^{\prime} -\tau)]$ and $J(\omega_k)$ represents the spectral density of an electromagnetic field inside the cavity. Lets assume that the spectral density of the electromagnetic field inside the cavity has Lorentzian form
\begin{equation}
J(\omega_k)=\frac{1}{2 \pi}\frac{\gamma \lambda^{2}}{(\omega_n-\omega_k)^{2}+\lambda^{2}},
\end{equation}
where $\omega_k$ shows the frequency of $k$-th cavity mode and $\omega_n$ is the center frequency of the cavity modes. $\lambda$ is the spectral width of the coupling, which is related to the cavity correlation time $\tau_c$ through $\tau_c=1/\lambda$. The parameter $\gamma$ is related to the scale time $\tau_c$, during which the system changes, through $\tau_s=1/\gamma$.  Strong and weak coupling regimes can be distinguished by comparing both the $\tau_c$ and $\tau_s$ time scales. When $\tau_s > 2 \tau_c$ ($\gamma<\lambda/2$), we have the weak coupling regime and the dynamics is Markovian \cite{52,53}. While the strong coupling regime corresponds to the case in which $\tau_s<2\tau_c$ ($\gamma>\lambda/2$) where the dynamics is non-Markovian. In the continuous limit ($\tau \longrightarrow \infty$) when $t>t^{\prime}$, the memory kernel can be written as
\begin{equation}\label{ss15}
F(t,t^{\prime})=\frac{\gamma \lambda}{4}\cosh\left[\theta(t-t^{\prime})  \right]e^{\bar{\lambda}(t-t^{\prime})},
\end{equation}
where $\bar{\lambda}=\lambda-i \Delta$ and $\theta = \beta(\bar{\lambda}+i \omega_0)$. By putting Eq. (\ref{ss15}) to Eq. (\ref{ss11}) and using Bromwich integral formula, the  probability amplitudes  $\tilde{A}(t)$ can be obtained as follows
\begin{align}\label{ss16}
\tilde{A}(t)&=\frac{(q_1+u_+)(q_1+u_-)}{(q_1-q_2)(q_1-q_3)}e^{q_1\gamma t} \nonumber \\
&-\frac{(q_2+u_+)(q_2+u_-)}{(q_1-q_2)(q_2-q_3)}e^{q_2\gamma t} \nonumber \\
&+\frac{(q_3+u_+)(q_3+u_-)}{(q_1-q_3)(q_2-q_3)}e^{q_3\gamma t},
\end{align}
where $q_i$'s ($i=1,2,3$) meet the following cubic equation as
\begin{equation}
q^{3}+2(y_1-iy_3)q^2+(u_+u_-+\frac{y_1}{4})q+\frac{y_1(y_1-iy_3)}{4}=0,
\end{equation}
where $y_1=\lambda/\gamma$, $y_2=\omega_0/\gamma$, $y_3=(\omega_0-\omega_n) /\gamma$, and $u_\pm=(1\pm\beta)\pm i\beta y_2-i(1\pm \beta)y_3$.
Using Eq. (\ref{ss16}), the evolved density matrix can be obtained as follows
\begin{equation}\label{ere}
 \rho(t)=\left(
\begin{array}{cc}
 \vert c_1 \vert ^{2} \vert A(t) \vert ^{2}  & c_1c^{*}_2 A(t) \\
 c_2c^{*}_1 A^{*}(t) &  1-\vert c_1 \vert ^{2} \vert A(t) \vert ^{2}  \\
\end{array}
\right).
\end{equation}

\section{QSL time for the model}\label{quantum}
In this section, we first review one of the comprehensive criteria for determining QSL time for open quantum systems \cite{28} and then the QSL time is studied for moving qubit inside a leaky cavity. The dynamics of an open quantum system is described via the following master equation
\begin{equation}
\dot{\rho}_{t}=\mathcal{L}_{t}\left(\rho_{t}\right),
\end{equation}
where $\rho_t$ is the evolved state of the system at time $t$ and $\mathcal{L}$ is the positive generator. QSL time is defined as the shortest time required to evolved from $\rho_\tau$ to  $\rho_{\tau+\tau_D}$, where $\tau$ is the initial time and $\tau_D$ is driving time. In Ref.\cite{28}, the authors have used relative purity to calculate QSL time for open quantum systems. They have shown that the relative purity between the state at initial time $\tau$ and the target state at time $\tau+\tau_D$ can be written as
\begin{equation}
f\left(\tau+\tau_{D}\right)=\frac{\operatorname{tr}\left(\rho_{\tau} \rho_{\tau+\tau_{D}}\right)}{\operatorname{tr}\left(\rho_{\tau}^{2}\right)}.
\end{equation}
By following the method outlined in Ref.\cite{28}, the ML bound for QSL time  can be obtained as follows
\begin{equation}\label{e1}
\tau \geq \frac{\left|f\left(\tau+\tau_{D}\right)-1\right| t r\left(\rho_{\tau}^{2}\right)}{\overline{\sum_{i=1}^{n} \kappa_{i} \varrho_{i}}},
\end{equation}
where $\varrho_i$ and $\kappa_i$ are the singular values of $\rho_{\tau}$  and $\mathcal{L}_{t}(\rho_{t})$, respectively, and $\overline{\square}=\frac{1}{\tau_{D}} \int_{\tau}^{\tau + \tau_{D}} \square dt$. The MT bound of QSL time can also be obtained similarly as follows
 \begin{equation}\label{e2}
\tau \geq \frac{\left|f\left(\tau+\tau_{D}\right)-1\right| \operatorname{tr}\left(\rho_{\tau}^{2}\right)}{\overline{\sqrt{\sum_{i=1}^{n} \kappa_{i}^{2}}}}.
\end{equation}
By merging these two bound for QSL time, a comprehensive bound can be achieved as follows
\begin{equation}
\tau_{Q S L}=\max \left\{\frac{1}{\overline{\sum_{i=1}^{n} \kappa_{i} \varrho_{i}}}, \frac{1}{\overline{\sqrt{\sum_{i=1}^{n} \kappa_{i}^{2}}}}\right\} \times\left|f\left(\tau+\tau_{D}\right)-1\right| \operatorname{tr}\left(\rho_{\tau}^{2}\right).
\end{equation}
It has also shown that the ML bound in Eq. (\ref{e1}) is tighter than the MT bound in Eq. (\ref{e2}) \cite{28}. The QSL time is inversely related to the speed of evolution in such a way that with increasing QSL time, the speed of evolution decreases and vice versa. In this work, we consider the maximally coherent initial state. This means that we put $c_1=c_2=1/\sqrt{2}$ in Eq. (\ref{initial}). So, the evolved density matrix has the following form
\begin{equation}\label{ere}
 \rho(t)=\frac{1}{2}\left(
\begin{array}{cc}
  \vert A(t) \vert ^{2}  &  A(t) \\
 A^{*}(t) &  2-\vert c_1 \vert ^{2} \vert A(t) \vert ^{2}  \\
\end{array}
\right).
\end{equation}
Let us calculate the singular values of $\rho_\tau$ and $\mathcal{L}_t (\rho_t)$, respectively. For $\rho_\tau$, the singular values are
\begin{equation}
\varrho _{1,2}=\frac{1}{2}\mp\frac{1}{2} \sqrt{\left| A(\tau)\right| ^4-\left| A(\tau)\right| ^2+1}.
\end{equation}
The singular values $\kappa_i$ of $\mathcal{L}_t (\rho_t)$ can be obtained as
\begin{small}
\begin{equation}
\kappa_{1}=\kappa_{2}=\frac{1}{2} \sqrt{\omega _0^2 | A(t)| ^2+| \dot{A}(t)| ^2-i \omega _0 A(t) \dot{A}(t)^*+i \omega _0 A(t)^* \dot{A}(t)+4 A(t)^2 \dot{A}(t)^2}.
\end{equation}
\end{small}
The QSL time for the qubit to evolve from $\rho_\tau$ to $\rho_{\tau+\tau_D}$ in this model can be obtained as
\begin{equation}\label{qslfinal}
\tau_{Q S L}^{A D}=\frac{\left|f\left(\tau+\tau_{D}\right)-1\right| \operatorname{tr}\left(\rho_{\tau}^{2}\right)}{\frac{1}{\tau_{D}} \int_{\tau}^{\tau+\tau_{D}}\left(\varrho_{1} \kappa_{1}+\varrho_{2} \kappa_{2}\right) d t},
\end{equation}
where
 \begin{align}
\left|f\left(\tau+\tau_{D}\right)-1\right| tr\left(\rho_{\tau}^{2}\right)&=\vert- 2 \vert A(\tau) \vert ^4  + 2(\vert A(\tau) \vert^2 -1)\vert A(\tau+\tau_D) \vert^2 \nonumber \\
&+ e^{-i k \omega_0} A(\tau+\tau_D) A(\tau)^{\star} + e^{i k \omega_0} A(\tau+\tau_D)^{\star} A(\tau) \vert.
 \end{align}

\begin{figure}[H]
  \centering
  \includegraphics[width=0.80\textwidth]{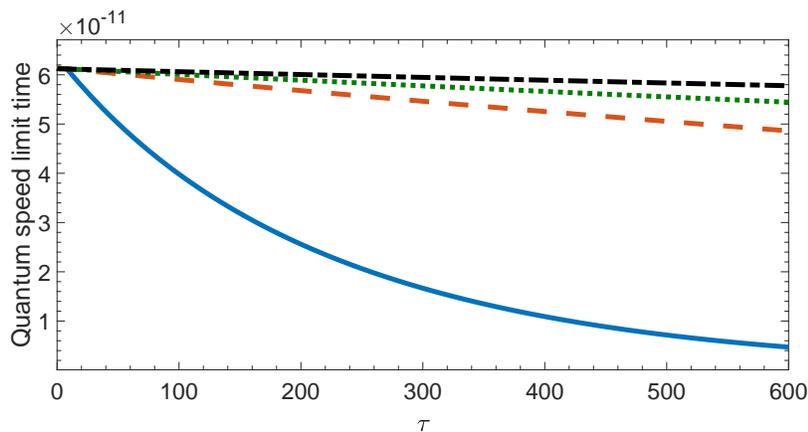}
\caption{QSL time for moving qubit inside leaky cavity versus initial time $\tau$ when $\lambda = 3\gamma_0$ and $\omega_0=\omega_n=1.53GHz$, for different values of the velocity of the moving qubit.  $ \beta=15 \times 10^{-9}$ (solid blue line),   $ \beta=50 \times 10^{-9}$ (dashed orange line),  $ \beta=70 \times 10^{-9}$ (dotted green
line),  $ \beta=100 \times 10^{-9}$ (dot-dashed black line).}
\label{figure2}
\end{figure}

In Fig. \ref{figure2}, the QSL time (\ref{qslfinal}) is plotted in terms of initial time $\tau$ for weak coupling regime ($\gamma<\lambda/2$). Also, different values of the velocity of moving qubit $\beta$ have been considered. In a weak coupling regime, the dynamics is Markovian. As can be seen, the QSL time increases with increasing the velocity of the moving qubit. It can also be seen that with increasing qubit velocity, QSL time almost reaches a constant value and it leads to a uniform speed for the dynamics. In fact, by increasing the velocity of the qubit inside the leaky cavity, coherence is protected, so the speed of evolution decreases. It can be stated that by increasing the qubit velocity inside the leaky cavity, the single-qubit system will be more stable and the process of losing coherence will be slower.

\begin{figure}[H]
  \centering
  \includegraphics[width=0.80\textwidth]{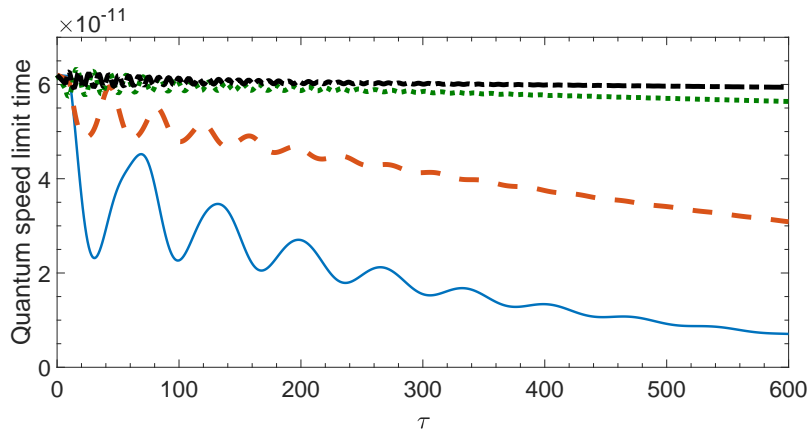}
\caption{QSL time for moving qubit inside a leaky cavity versus initial time $\tau$ when $\lambda = 0.01\gamma_0$ and $\omega_0=\omega_n=1.53GHz$, for different values of velocity of the moving qubit.  $ \beta=0.05 \times 10^{-9}$ (solid blue line),   $ \beta=0.1 \times 10^{-9}$ (dashed orange line),  $ \beta=0.3 \times 10^{-9}$ (dotted green
line),   $ \beta=0.5 \times 10^{-9}$ (dot-dashed black line).}
\label{figure3}
\end{figure}
In Fig. \ref{figure3}, the QSL time is plotted as a function of initial time for strong coupling regime ($\gamma>\lambda/2$). As mentioned before, in a strong coupling regime the dynamics is non-Markovian and information flow back from the environment to the system.  In this figure, the QSL time is plotted for different values of the velocity of moving qubit $\beta$. As can be seen, the QSL time has an oscillating behavior and decreases for small values of the velocity of the qubit. It is also observed that the QSL time increases with increasing the velocity of the qubit inside the leaky cavity. One can also be seen that with increasing qubit speed, QSL time almost reaches a constant value and therefore leads to a uniform evolution speed for the mentioned open system.

\section{Conclusion}\label{con}
In the quantum world, we are mostly dealing with open quantum systems. Due to the importance of open quantum systems, the study of the dynamics and properties of these systems is of particular interest. One of the features that can be considered in both closed and open quantum systems is the QSL time. In this work, a structure is considered as an open quantum system in which a single-qubit moves inside a leaky cavity. The motivation for choosing such a structure is that in this structure, as a result of the movement of single-qubit with high speed, the degree of coherence of the initial state remains more stable during the evolution. In this structure, both strong coupling and weak coupling of the system with cavity modes were considered. In a weak coupling regime, the dynamics is Markovian (memory-less) and information is leaked from the system to the environment monotonically and there is no flow back of information. While in a strong coupling regime, the dynamic is non-Markovian (with memory) and information flow back from the environment to the system during the evolution.

In this work, the QSL time was studied for the mentioned model, i.e., the moving qubit inside a leaky cavity. In the case of a weak coupling regime, it was observed that the QSL time decreases monotonically with increasing initial time $\tau$. It was also observed that with the increasing velocity of the qubit inside the leaky cavity, the QSL time increases. Due to the inverse relationship between QSL time and speed of the evolution, it can be said that by increasing the qubit velocity inside the leaky cavity, the speed of evolution decreases and the system is more stable. Of course, such a result was predictable due to the direct relation between the coherence of the initial state and QSL time.
In the case of a strong coupling regime, it was observed that the QSL time decreases with an oscillating behavior. It also observed that the QSL time increases with increasing the velocity of qubit inside the leaky cavity.
In summary, one can be noted that in the mentioned model, for both strong and weak coupling regimes, increasing the velocity of qubit inside a leaky cavity leads to increasing QSL time and slowing down the evolution.

\section*{ORCID iDs}
Soroush Haseli \href{https://orcid.org/0000-0003-1031-4815}{https://orcid.org/0000-0003-1031-4815}\\
Hazhir Dolatkhah \href{https://orcid.org/0000-0002-2411-8690}{https://orcid.org/0000-0002-2411-8690}\\
Saeed Haddadi \href{https://orcid.org/0000-0002-1596-0763}{https://orcid.org/0000-0002-1596-0763}

\section*{Conflict of interest}
The authors declare no competing interests.

\end{document}